%% file: example_paper.tex
\documentclass[nohyperref]{article}

\usepackage{microtype}
\usepackage{graphicx}
\usepackage{subfigure}
\usepackage{booktabs} 

\usepackage[hyphens]{url}
\usepackage{hyperref}

\usepackage[accepted]{icml2022}

\usepackage{amsmath}
\usepackage{amssymb}
\usepackage{mathtools}
\usepackage{amsthm}

\usepackage[capitalize,noabbrev]{cleveref}

\theoremstyle{plain}

\theoremstyle{definition}

\theoremstyle{remark}

\usepackage[textsize=tiny]{todonotes}

\icmltitlerunning{The Edge of Disaster: A Battle Between Autonomous Racing and Safety}

\begin{document}

\twocolumn[
\icmltitle{The Edge of Disaster: A Battle Between Autonomous Racing and Safety}



\icmlsetsymbol{equal}{*}

\begin{icmlauthorlist}
\icmlauthor{Matthew Howe}{equal,adl}
\icmlauthor{James Bockman}{equal,adl}
\icmlauthor{Adrian Orenstein}{equal,adl}
\icmlauthor{Stefan Podgorski}{adl}
\icmlauthor{Sam Bahrami}{adl}
\icmlauthor{Ian Reid}{adl}

\end{icmlauthorlist}

\icmlaffiliation{adl}{Australian Institute for Machine Learning, University of Adelaide, Adelaide, Australia}

\icmlcorrespondingauthor{Matthew Howe}{matthew.howe@adelaide.edu.au}

\icmlkeywords{Machine Learning, ICML}

\vskip 0.3in

]
\hspace*{20pt}
\includegraphics[width=0.95\textwidth]{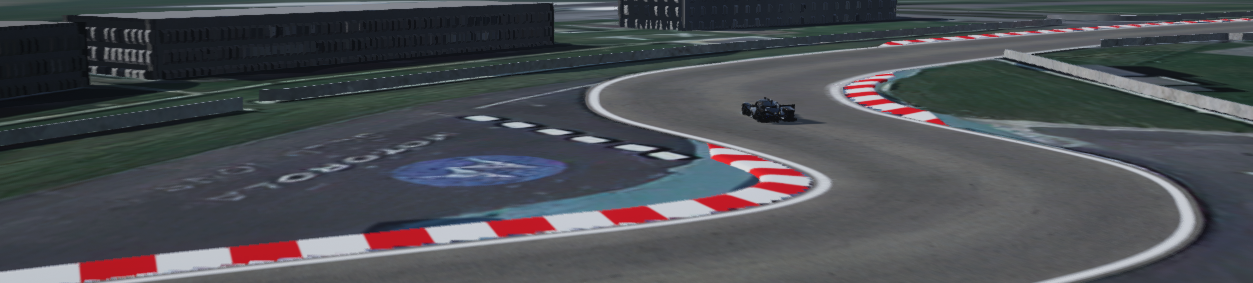}


\printAffiliationsAndNotice{\icmlEqualContribution} 

\begin{abstract}
Autonomous racing represents a uniquely challenging control environment where agents must act while on the limits of a vehicle's capability in order to set competitive lap times.
This places the agent on a knife's edge, with a very small margin between success and loss of control.
Pushing towards this limit leads to a practical tension: we want agents to explore the limitations of vehicle control to maximise speed, but inadvertently going past that limit and losing control can cause irreparable damage to the vehicle itself.
We provide a model predictive control (MPC) baseline that is able to, in a single lap, safely adapt to an unseen racetrack and achieve competitive lap times.
Our approaches efficacy is demonstrated in simulation using the Learn To Race Challenge's environment and metrics. \cite{herman2021learn}
\end{abstract}

\section{Introduction}
Racing is a hallmark of automotive innovation.
Competitors and manufactures alike push the very limits of physical capability and design to edge out the slightest of advantages over their opponents.
Gains at the top level of motorsport are often realised in hundredths of seconds, but can have significant impact on the drivability, safety and efficiency of road cars.

\newpage
\vspace*{90pt}

Mercedes-Benz lists a slew of technical innovations used in their road cars that are directly a result of their Formula One participation. \cite{colquhuon_2021}
In this spirit, racing leagues for vehicles piloted by autonomous systems have started across the world. \cite{DBLP:journals/corr/abs-1911-01562,DBLP:journals/corr/abs-1901-08567,DBLP:journals/corr/abs-1806-00678,DBLP:journals/corr/abs-1908-08031,wischnewski2022indy}
Competitors vie to create the best platforms and automated control in the never ending quest for faster lap times.
However, as motorsport hall of fame inductee Sir Stirling Moss advised, \textit{"To achieve anything in this game you must be prepared to dabble in the boundary of disaster."}.
This presents a natural conflict. 
In our quest for speed, the edge of disaster is flirted with and often indulged, causing catastrophic damage to valuable racing platforms. \cite{tingwall_2021,fingas_2021}
It is therefore desirable to design control systems that, much like a human drivers, progressively explore the physical limits of the platform in a way that minimises risk.
To this end, the Learn To Race Challenge encourages competitors to not only develop autonomous control capable of racing pace, but penalises systems for unsafe behaviour. \cite{herman2021learn}
In addition to a focus on safety, the challenge limits available information from the environment and tests the few-shot learning capability of systems.
Observations from the sim at evaluation time are images from a predefined RGB camera system and the current vehicle speed.
This requires systems to infer spatial information about the vehicles position exclusively from visual information.
Systems adaptability to new tracks is evaluated by providing a one hour practice session where the system is able to drive around an unseen track.
During this practice window, safety infraction penalties are accumulated while any automatic adaptions are made by the autonomous system to adapt to the new environment.
Three factors: safety, limited observation and few-shot generalisation provide a uniquely challenging environment to develop autonomous control.

Although the challenge encourages submissions to explore systems leveraging safe reinforcement learning our initial efforts where directed toward producing a baseline of more tradition approaches to autonomous control which could be built upon.
Unfortunately the majority of these techniques rely on knowledge of the track layout and accurate localisation around the circuit; two things this challenge restricts.
As a result we provide details of the adaption of several control solutions tailored to the restrictions specific to the 2022 Learn To Race Challenge.
These include a naive Follow-The-Gap driver, a model predictive controller (MPC) that uses local perception and a MPC based system that uses localisation and mapping to aid control (MPC++).
Our baseline MPC solution achieved competitive results in the Learn To Race Challenge.
The implementation of MPC++ was finalised after the conclusion of the submission deadline and therefore was not officially in contention.

\section{Related work}\label{sec:related-work}
Our approach to solving the autonomous racing problem was to follow the simple and effective autonomous driving pipeline of perception, mapping, localisation, planning, and control.
Initially we worked at solving and testing each of these problems independently to build an entire working system.
With this modular approach, we could improve and iterate on each subsystem without having to rebuild the entire system, like in end-to-end control systems.

The perception module was inspired by the work of \cite{simtoreal}, which presented the benefits of abstracting perceptions from raw sensor inputs to a simplified and scene independent format.
Our control systems were inspired by \cite{mpcminicars,AMZdriverless} which utilise model predictive controllers for autonomous racing.
Liniger's MPC has been used to both control vehicle behaviour at the limits of grip at 1/43rd scale and in Formula SAE races. \cite{Liniger_2014,AMZdriverless}
Augmenting these controllers with learning has been explored in \cite{learningmpc}.
They achieve improved dynamic control using a machine learning model to predict refinements to MPC output.
In the work explored the vehicle's location is known or there are discrete visual markers present---such as traffic cones---that enable robust point matching; producing high accuracy localisation.

In this challenge we must handle vehicle localisation with sparse visual features, see Figure \ref{fig:visualfeatures}, and construct our own map of the circuit.
This leads to more uncertainty in planning and control than these systems are capable of dealing with in their current form.
Ultimately, a successful solution to this challenge must be built with this uncertainty in localisation, planning and control in mind.

\section{Methodology}
In this section we outline the inner workings of our autonomous racing agent.
We first outline a naive Follow-The-Gap agent in Section \ref{sec:followthegap} which achieved a top 10 position in stage-one of the competition.
Then we provided details of a MPC based agent in Section \ref{sec:MPC} which achieved competitive results in stage-two of the 2022 Learn To Race Challenge.
Additionally, a final system denoted as MPC++ which was not submitted for evaluation is presented in Section \ref{sec:MPC++}.

\subsection{Follow-The-Gap Driver}\label{sec:followthegap}
\begin{figure*}
    \centering
    \begin{tabular}{cll}
    \includegraphics[width=0.36\textwidth]{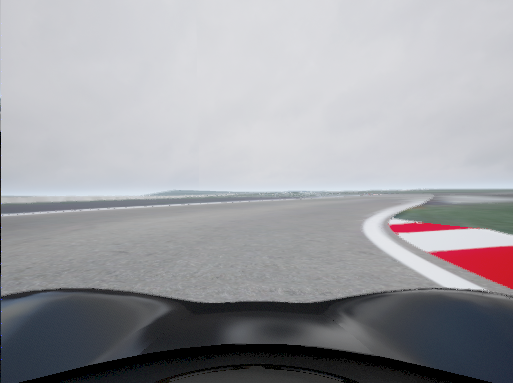} &
    \includegraphics[width=0.3\textwidth]{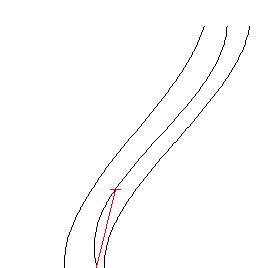} &
    \includegraphics[width=0.3\textwidth]{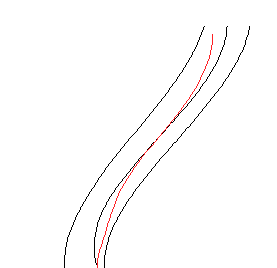} \\
    (a) & \hspace{40pt} (b) & \hspace{40pt} (c)    
    \end{tabular}
    \caption{(a) Input camera image (b) Extracted track limits (black) and predicted path (red) from Follow-The-Gap driver control (c) Extracted track limits (black) and predicted path (red) from MPC control}
    \label{fig:control}
\end{figure*}

As an initial pilot test for control, based on our road segmentation model, we implemented a naive Follow-The-Gap driver.
This controller uses the driveable area segmentation mask of the road in front of the vehicle projected onto the ground plane to plan a straight line trajectory while maintaining a safe distance from all track limits, see Figure \ref{fig:control}(b).
Acceleration and steering inputs to the vehicle are set proportionately to the length and angle of the trajectory respectively.
The constants of proportionality used for this driver where manually tuned until desirable driving behaviour was achieved.
To extract a representation for the track limits in front of the vehicle we developed a Perception module.

\begin{figure}
    \centering
    \resizebox{0.49\textwidth}{!}{
        \begin{tabular}{cc}
            \includegraphics[width=0.24\textwidth]{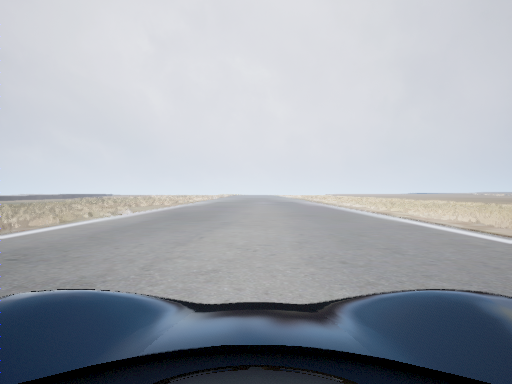}&
            \includegraphics[width=0.24\textwidth]{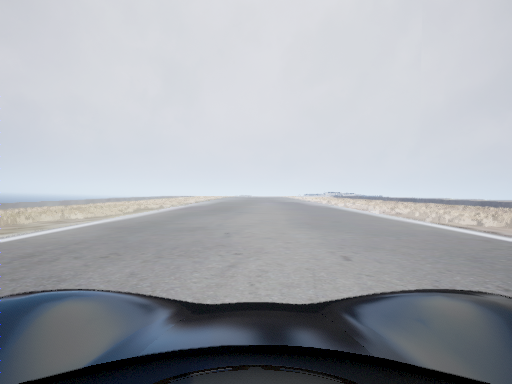}\\
            (a)&(b)\\
            \includegraphics[width=0.24\textwidth]{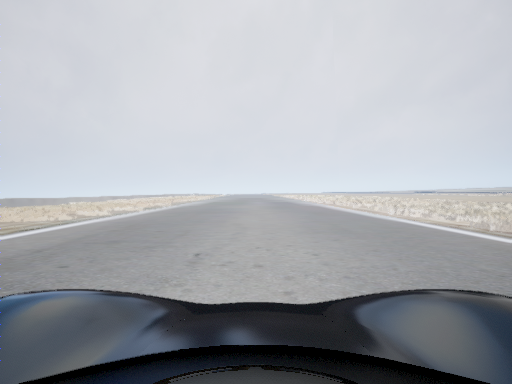}&
            \includegraphics[width=0.24\textwidth]{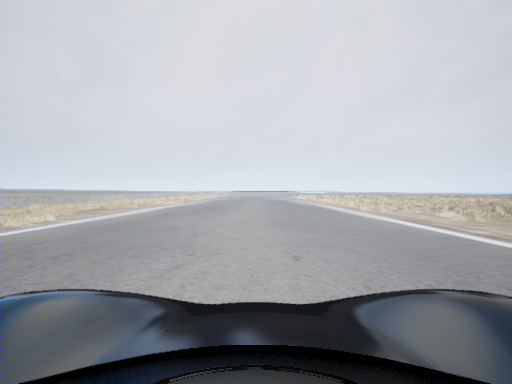}\\
            (c)&(d)
        \end{tabular}
    }
    \caption{Image captures showing the visual features around Anglesey from the vehicles front camera (a) after turn 1 (b) after turn 2 (c) after turn 5 (d) before turn 6}
    \label{fig:visualfeatures}
\end{figure}

\subsubsection{Perception}\label{sec:perception}
Perception is responsible for processing camera images into a representation of the local limits of the road.
This abstraction to track limits is used by all other modules.
To achieve this a combination of deep learning and handcrafted post-processing is used.
Using ground truth segmentation maps provided by the simulation two different segmentation models where trained.
For the Follow-The-Gap and MPC solutions a Feature Pyramid Network \cite{FPN2017} with an EfficentNet \cite{tan2019efficientnet} backbone was used.
In MPC++ we substituted this for Deeplabv3+ \cite{DBLP:journals/corr/abs-1802-02611} with a ResNet-18 \cite{DBLP:journals/corr/HeZRS15} encoder.
Both where trained to classify whether a given image pixel belongs to the drive-able portion of the race track or not.
Modelling is preformed using PyTorch \cite{DBLP:journals/corr/abs-1912-01703} on a bespoke dataset we collected from the simulation environment.
For more details on model training see Appendix \ref{app:modelling}
This produces a binary mask representing the drive-able region in front of the vehicle.
Three sets of points are extracted from the mask that are representative of the left and right track limits in addition to a set of points along the centre of the drivable area.
For the left and right track limits the left and right most pixels classified as drive-able for each row in the mask are selected.
Similarly, centreline points are extracted by selecting the middle pixel in a given row classified as drivable.

Track limits are then projected from image coordinates $(u,v)$ onto the ego-centric ground-plane $(x_{\text{ego}}, y_{\text{ego}})$, where $x_{\text{ego}}$ and $y_{\text{ego}}$ represent lateral and longitudinal displacement from the vehicle centre respectively.
We utilise a homography matrix to map points from camera to ego-centric vehicle coordinate frames:
\begin{equation*}
    s
    \begin{bmatrix}
        x_{\text{ego}}\\
        y_{\text{ego}}\\
        1
        \end{bmatrix}
        =
        H_{3\times3}
        \times
        \begin{bmatrix}
        u\\
        v\\
        1
    \end{bmatrix}
    \label{eq:homo}
\end{equation*}
To estimate the homography matrix we must have a minimum of four corresponding points between the camera and ground plane.
Using the known camera parameters from the simulation environment the intrinsic and extrinsic camera calibrations can be found:
\begin{equation*}
    s
    \begin{bmatrix}
    u\\
    v\\
    1
    \end{bmatrix}
    =
    \begin{bmatrix}
    f & 0 & c_x\\
    0 & f & c_y \\
    0 & 0 & 1
    \end{bmatrix}
    \times
    \begin{bmatrix}
    r_{11} & r_{12} & r_{13} & t_1\\
    r_{21}& r_{22} & r_{23} & t_2\\
    r_{31} & r_{32} & r_{33}& t_3
    \end{bmatrix}
    \begin{bmatrix}
    X\\
    Y\\
    Z \\
    1
    \end{bmatrix}
\label{eq:camera}
\end{equation*}
where $c_x$ and $c_y$ are the centre positions of the image plane in pixels.
The focal lengths $f_x$ and $f_y$ are calculated using the field of view and the image size:
\begin{equation*}
    f = \frac{w}{2\times \tan(FOV_h/2)}
\label{eq:focallength}
\end{equation*}
where $w$ is the image width in pixels and $FOV_h$ is the horizontal field of view.

\begin{figure}
    \centering
    \includegraphics[scale=0.22]{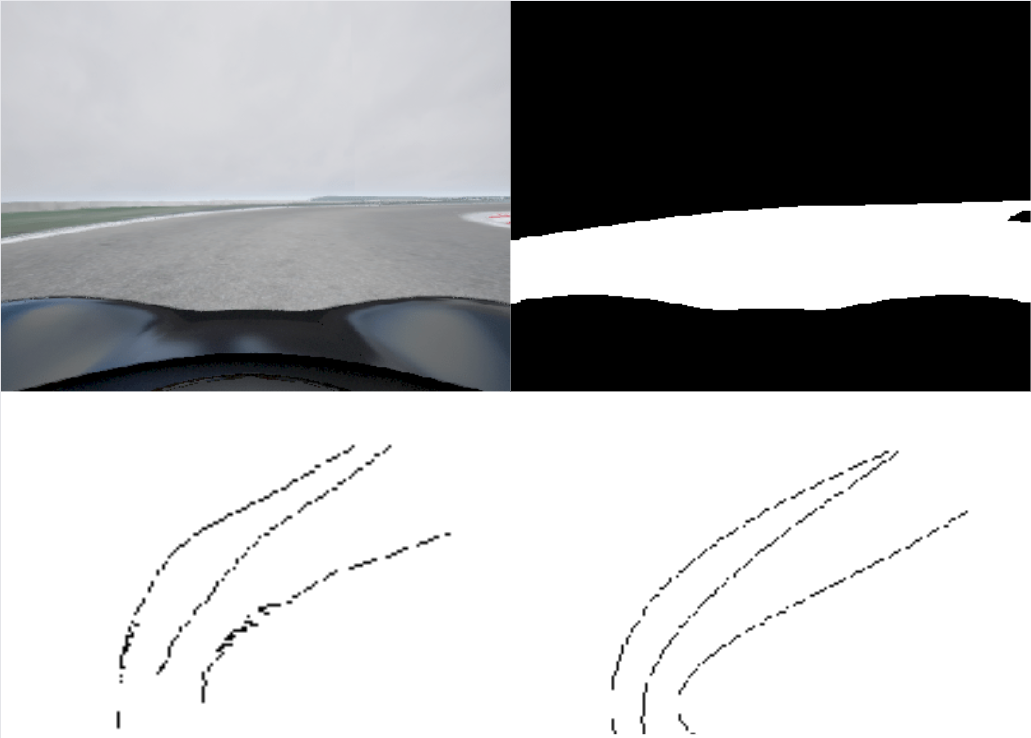}
    \caption{An example of the perception module processing a camera image into a road segmentation mask, then extracting track points from the mask and smoothing the extracted track points.}
    \label{fig:perception}
\end{figure}

Points extracted and transformed using this technique are variable in both density and number. 
For example, track limits closer to the camera have more points compared to those further away.
This method also assumes that the ground and camera plane remain fixed relative to one another.
Additionally, representing track curves on a squared grid results in jagged edges in extracted lines.
These factors can mean even a perfect track segmentation can have errors induced from vehicle movement and track elevation changes leading to problems further down the pipeline.
To remedy some of these factors, a third order polynomial is fit to each set of extracted points separately and new points are recalculated from the resulting equation.
Figure \ref{fig:perception} shows an example of how the perception system translates images into a representation of the local track.
These three sets of points---left, right and central---are the representations used by the other subsystems.

\subsection{MPC}\label{sec:MPC}
For this solution, the Follow-The-Gap controller is replaced with one using MPC.
Here we detail how the control problem can be framed as an optimisation problem and the various modifications required to enable operation with local perceptions.
Herein we refer to the part of the system responsible for translating track-limit perceptions to steering and acceleration inputs.

\subsubsection{Control}\label{sec:control}
We modify the off the shelf MPC controller \cite{mpc} which utilises the quadratic program solver \cite{osqp}.
This takes a representation of driveable area in ego-centric vehicle coordinates from the front cameras (Sec \ref{sec:perception}) i.e. the local road view, and outputs a series of control inputs.
The resulting trajectory minimises local segment time and deviation from a reference path while abiding by safety and vehicle constraints. 

\newpage

As track limit perceptions are in a ego-centric coordinate frame we convert these to a spatial representation.
The spacial representation used contains the curvature, $\kappa = 1/r$, of the path, distance between each of the points, and the track width.
Using this information, a path that minimises heading error, displacement error, and time taken to travel between points is found.
Each error term has an associated weight which is used to tune the vehicle's behaviour.
For example, by weighting time highly paths which do not follow the reference path exactly become permissible.
This allows deviation from the centre of the racetrack, prioritising paths that are more time efficient and therefore faster.
Alternatively, if we increase the weighs for displacement and heading error the vehicle will follow the detected centreline closely, at the expense of speed.

Calculating control can be broken into two main steps:
First, a desired speed profile is calculated which constraints the maximum and minimum acceleration and lateral acceleration based on track curvature.
Second, vehicle yaw and velocity is optimised for each reference track point. 
The yaw error, track error and time are constrained such that only physically possible solutions are produced.
Formally, the optimal control problem is defined as follows:
\begin{equation*} \label{mpc}
\begin{aligned}
    \underset{v, \psi}{\text{minimize}}
        && \gamma_{p}\sum_{i=0}^{i=n} \vert\vert p_{i} - \text{ref\_}p_{i} \vert\vert + \gamma_{t} T \\
        && + \gamma_{v}\vert\vert \Delta v \vert\vert + \gamma_{s}\vert\vert \Delta s \vert\vert  \\
    \text{subject to}
        && v_{n} \leq v_{end}, \\
        && \vert p_{i} - \text{ref\_}p_{i} \vert \leq \frac{w}{2}, \\
        && v_{min} \leq v_{i} \leq v_{max}, \\
        && s_{min} \leq s_{i} \leq s_{max}, \\
        && v_0 = v, \\
        && \psi_0 = 0
\end{aligned}
\end{equation*}
where $\gamma_p$, $\gamma_t$, $\gamma_v$ and $\gamma_s$ are the weights for position error at each control step, overall time, magnitude of velocity change and magnitude of steering change respectively, $p$ is the vehicle position and ref$\_p$ is the desired position, $T$ is the time to complete the trajectory, $v$, $s$ and $\psi$ are the vehicle's velocity, steering angle and yaw, $w$ is the track width, $v_{end}$ is a maximum terminal velocity and $n$ is the number of steps in the time horizon.

Once the optimisation problem is solved steering angles are computed using a bicycle model:
\begin{equation*}
    \delta = arctan(\psi \times\text{wheelbase})
\end{equation*}
Acceleration inputs are formulated as the difference between the desired and reference speed:
\begin{equation*}
    a = \frac{v_{ref} - v}{\beta}   
\end{equation*}
where $\beta$ can be changed to modify how aggressively acceleration is changed. 
Figure \ref{fig:control} shows an example of the vehicle trajectories calculated using the Follow-The-Gap and MPC based methods.

\subsection{MPC++}\label{sec:MPC++}
This version of our system enables control to take into account larger context than what is available from local perceptions.
Steering inputs remain entirely dependant on the local track limit perception.
However, velocities can now be set with a look ahead much further than is often available locally and in a way that is robust to potentially erroneous perceptions.
To do this we add two additional modules responsible for localisation and mapping.
Mapping is done in the one hour practice session available prior to racing.
The resultant map is then used to localise the vehicle while racing.
Once vehicle position is approximately known control of velocity is ceded to references calculated from the map.

\begin{figure*}
    \centering
    \resizebox{\textwidth}{!}{
        \begin{tabular}{c}
            \includegraphics[width=0.5\textwidth]{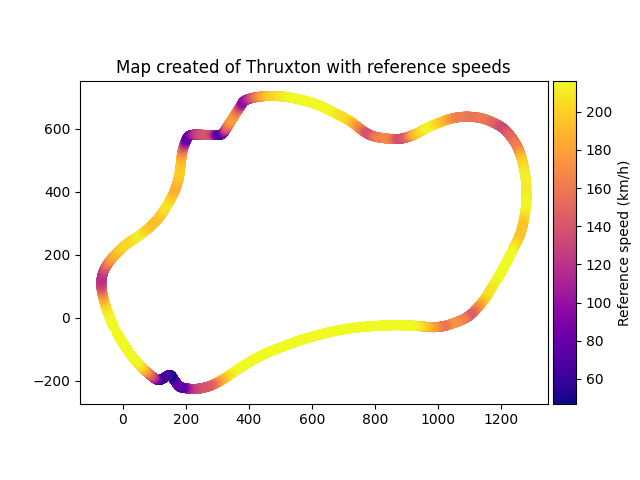}
            \includegraphics[width=0.5\textwidth]{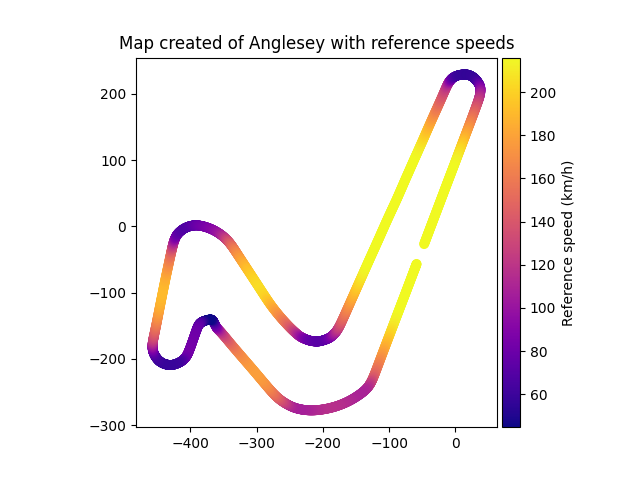}
        \end{tabular}
    }
    \caption{Mapping results from our agent at training time with computed max reference speeds}
    \label{fig:maps}
\end{figure*}

\subsubsection{Mapping}\label{sec:mapping}
Our agent has no knowledge of circuit layout prior to setting a lap time.
A mapping algorithm is used to create a representation of the track which has two key benefits.
It enables the location of the vehicle within the map to be estimated using the Localisation module, detailed in Section \ref{sec:localisation}, and allows for pre-computation of reference velocities informed by the global structure of the circuit.
In cases where the Perception module's line-of-sight is limited or observations are of low quality, these reference velocities provide an alternative signal to set the vehicle's speed.

To create this map, during practice, the MPC controller with conservative control constrains and a speed limit of $20\text{ms}^{-1}$ is used to circumnavigate the track.
These control settings are set such that the agent will safely complete a full lap and move slowly enough that detailed track limit points are able to be collected.
Points used to describe the track limits can be gathered over the course of one or more laps.
This results in two collections of points corresponding to the outside and inside limits of the circuit.
As these points are collected via the perception module, overlapping and duplicate points are appended sequentially; resulting in a out of order set.
To recover ordering between sequential points the Pyconcorde traveling salesperson problem solver was used. \cite{pyconcorde}
The now ordered points are smoothed using a Savitzky–Golay filter. \cite{Savitzky1964SmoothingAD}
From the pair of track limits a centreline is also interpolated.
In our case, the pair of track limit's detection spacing varies over the course of collection, so bipartite matching of inside and outside points was not possible.
Instead, track limit points are matched to each other by finding their closest point on the opposing set of points.
Once matched, the pair's midpoint is used as a centreline point.
Using the centreline a series of reference velocities for the circuit are calculated which take into account the maximum lateral acceleration and track curvature.
To do this we used the velocity profile optimisation, discussed in Section \ref{sec:control}, to calculate the reference velocity assigned to each centreline point.
Velocity profile optimisation takes into account the vehicle's maximum lateral acceleration, acceleration and braking capabilities.
The results of this processes are shown in Figure \ref{fig:maps}.

\subsubsection{Localisation}\label{sec:localisation}
In order to utilise the reference speeds calculated for each position around the circuit, we need to know where the vehicle is.
To estimate the vehicle's track position our localisation module uses track limit detections and a particle filter.
More specifically, particles begin uniformly spread through the previously computed circuit map, see Section \ref{sec:mapping}.
At each time step, particles positions are progressed using vehicle control input, a dynamics model and random noise:
\begin{equation*}
    \begin{aligned}
        x_{t+1} &= x_{t} + \Dot{x} \times dt, \\
      \Dot{x} &=
        \begin{bmatrix}
            v \times cos(\psi) \\
            v \times sin(\psi) \\
            v \times tan(\delta) / b
        \end{bmatrix}
        \label{eq:bicyclemodel}
    \end{aligned}
\end{equation*}
where $x$ and $b$ are the vehicle's pose and wheelbase respectively.

Observations in the form of track limit detections from the Perception module are transformed to each particle's location.
Using a k-d tree particle's closest points in the map are found and a local section of the map's track limits in front of the particle are extracted.
These points are lifted from the map sequentially under the assumption that both the observed track limits and map points are equally spaced.
In practice this is not always the case.
We ensure the observed track points are over sampled and then uniformly drop points until the observation and map densities are equivalent.
The $\ell2$ error between observed detections and map points is evaluated and used to score and remove particles.

Particles are removed from the filter if they: they fall out from within the track limits of if the track limit matching score is too low.
New particles are generated by resampling particles in the filter.
Each particle is sampled with a probability proportional to their associated score.
A particle's score is calculated by normalising the value of a normal distribution's probability density function evaluated at the observation error, $E\sim N(0, \sigma)$.
This manifests as particles with no error---at the mean of the distribution--being assigned the highest score of 1, with scores decreasing as error increases.
The rate of score decrease is negatively correlated with $\sigma$.
This leads to low observation error particles being resampled more frequently than their high error counterparts.
Additionally, each particle's score is used to weight its overall contribution to the estimated position of the vehicle.
These two factors enable the filter to progressively refine the collection of particles it maintains; converging to an estimate of the vehicles location on the circuit.
To detect when the filter has \textit{converged} we use a threshold on the maximum distance a single particle can be away from the current estimated position.
Once all particles are within this distance we consider the filter to be \textit{converged}.
Once converged, the estimated position can be used to look-up pre-calculated reference velocities from the map.
These reference velocities enable vehicle control to preempt the circuit's profile further into the future than what is visually available from its cameras.

\section{Evaluation}
To compare different control systems lap times set on three different racetracks where used.
Thruxton and Anglesey National where provided by the competition organisers for local testing and development.
Examples of the track's layouts can be seen in Figure \ref{fig:maps}.
Las Vegas Outfield North was tested on via submission to an evaluation server.
Laps are timed with the vehicle starting on the start-finish line accelerating from a stand still.

\section{Results}
Evaluation of each outlined system of control are shown in Table \ref{tab:lap-times}.
All control systems explored in this work recorded no safety infractions.
Due to the competition shutting its evaluation servers we where unable to evaluate MPC++ on the Las Vegas racetrack. 

\begin{table}[h]
    \centering
    \resizebox{0.48\textwidth}{!}{
        \begin{tabular}{@{}lccc@{}}
            \toprule
                                 & \multicolumn{3}{c}{\textbf{Lap Times (m:s:ms)}}            \\
            \textbf{Driver}  & \textbf{Thruxton} & \textbf{Anglesey} & \textbf{Las Vegas} \\ \midrule
            Follow-The-Gap    & 2:13:835          & 1:20:187          & 1:45:388           \\
            MPC                  & 1:55:708          & 0:53:837          & \textbf{1:19:148}           \\
            MPC++                & 1:38:950          & 0:44:601          &       -             \\
            Tuned MPC++          & 1:29:516          & 0:43:204          &         -           \\
            Tuned MPC++ No Delay & \textbf{1:29:136} & \textbf{0:42:002} &        -            \\ \bottomrule
        \end{tabular}
    }
    \caption{
    Observed lap times around the three racetracks for different versions of our solution.
    Follow-The-Gap is provided as a naive control baseline which was submitted for stage one.
    MPC uses local perceptions only, submitted for stage two.
    MPC++ uses localised pace notes with parameters that work across all tracks.
    Tuned MPC++ is manually tuned to be faster around each track with settings that may not generalise to other tracks.
    No Delay indicates that the 10Hz input/observation frequency is turned off.
    }
    \label{tab:lap-times}
\end{table}

\section{Discussion}
Follow-The-Gap driver's naive control design is superseded, $-18.127$s ($13.5\%$), $-26.350$s ($32.9\%$) and $-26.24$s ($24.9\%$) on Thruxton, Anglesey and Las Vegas respectively, by the MPC control using local perceptions to guide its trajectory.
The difference in lap time between the two systems is observed to be correlated with the number of tight corners in a given racetrack.
In the collection of tracks used Thruxton contains more sweeping corners with very few that are tight and slow.
Conversely, Las Vegas contains several tight chicanes and hairpins.
Due to the Follow-The-Gap driver projecting a straight line through the segmentation mask, trajectories that deviate significantly from this template will be harder to achieve, Figure \ref{fig:control}.
In the case of a sweeping corner or long straight this would not be much of an issue.
However, as is the case of approximating a circle from a polygon, maneuvering through a tight hairpin requires many small piece-wise linear trajectories.
As the controller sets its speed according to the length of the trajectory this has the effect of jagged and slow control through such corners.
As the MPC controller is able to plan control inputs at multiple control points along a given section of track, steering and acceleration is much smoother in comparison.

\begin{figure*}
    \centering
    \begin{tabular}{ccc}
    \includegraphics[width=0.25\textwidth]{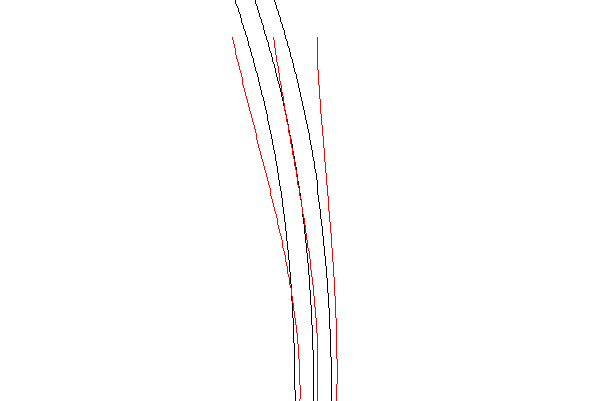} &
    \includegraphics[width=0.25\textwidth]{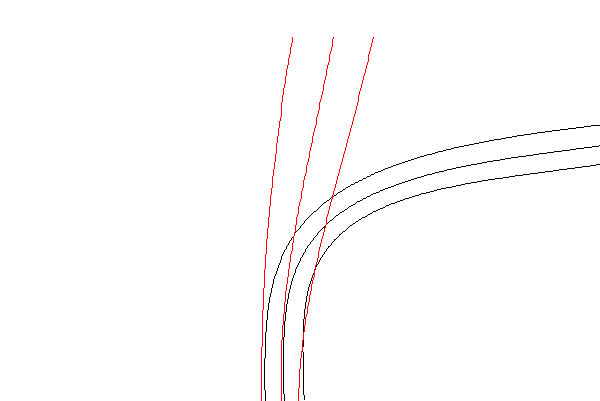} &
    \includegraphics[width=0.25\textwidth]{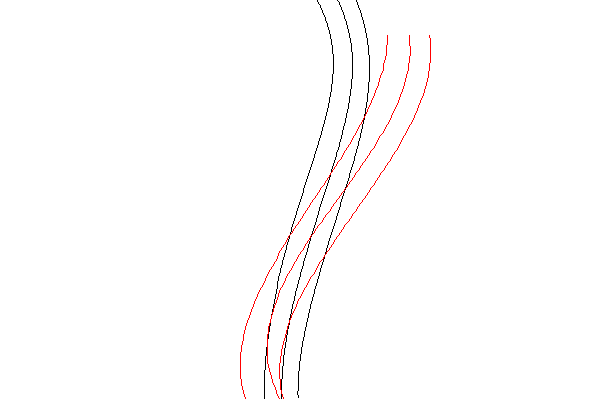} \\
    (a) & (b) & (c)    
    \end{tabular}
    \caption{Each image represents a pair of ground truth and estimated local track limits from the localisation module. Ground truth track limits are shown in black and estimations in red. (a) Localisation along a straight (b) Reaching a corner early (c) Error in yaw approximation}
    \label{fig:localisation}
\end{figure*}

Despite this improvement the local perception MPC controller suffers from a lack of foresight.
As it is restricted to using its camera to predict a path through the up coming section of road, it is unable to deal with situations such as a crest in the middle of a long straight.
Ideally in such a situation the vehicle would continue without adjusting its input with the knowledge that the straight continues over the crest.
The local MPC controller however needs to slow down in anticipation that there may be the sharpest turn on the track beyond the crest, as it has no understanding of its position and track geometry.
Quantitatively we demonstrate that the ability to \textit{look ahead} provides a significant improvement over the MPC using local perceptions.
MPC++ creates and localises to a map of the circuit.
By adding this an improvement of $-16.758$s on Thruxton and $-9.236$s Anglesey was observed; a reduction in lap time of $34.885\%$ and $35.586\%$ respectively when compared to follow-the-gap.

The results shown for MPC++ in Table \ref{tab:lap-times} are for general optimisation parameters that work well on both Thruxton and Anglesey.
To demonstrate the potential for an automated adaption system we manually tuned two separate sets of parameters specific to either Thruxton or Anglesey, denoted tuned MPC++.
With limited experimentation, further improvements of $-9.434$s on Thruxton and $-1.397$s on Anglesey were found.
Overall, a tuned MPC++ is capable of $44.7\%$ and $38.2\%$ faster laps around Thruxton and Angelesey respectively.
We think that by tuning MPC++ the behaviour around longer sweeping corners was improved while tighter corners were still challenging to perceive and control around.
Since Thruxton has more of the former we see a larger improvement in lap time relative to Anglesey.
Automation of this tuning via learning or model refinement is therefore recommend as a line of future enquiry.

\subsection{Runtime analysis}
Although we are limited to $10$Hz control and perception this assumes that our model and calculations are instantaneous.
This is because the observation delay is called as a sleep function in the evaluation server's code base.
For example if our code takes $0.04$s to run this would result in a control frequency of $7$Hz.
Without this delay control control could be calculated and applied at $25$Hz.
For this reason we focused on our code having the lowest run time possible as this has a great effect on the controllers ability to correct oversteer and poor control inputs.
One step of the MPC++ controller takes $0.025$s and has the ability to run at $40$Hz.
We found that since the sleep function runs on a separate thread we can run calculations during this time which means we were able to execute at $9.5$Hz with observation delay enabled.

\subsection{Limitations of our system}
We observed that the detected track limits from the perception system (Section \ref{sec:perception}) had large effects on the control output and performance of the vehicle.
Due to using a static transformation from camera to ground plane error is introduced as the projection changes when the vehicle squats, rolls, or dives.
This also occurs when the elevation of the track changes, which occurs on several occasions around the Thruxton circuit.
This could lead to the track limits spanning much further ahead of the vehicle than they are, resulting in the vehicle accelerating faster than what is safe.
Additionally, localisation accuracy is impacted as calculation of the error between the observed track limits and the map's is used in particle scoring and elimination.

Our system uses dead-reckoning and abstracted visuals for localisation.
This allows the system to go faster through straights by knowing the track ahead, see Figure \ref{fig:localisation}(a).
Since there we where unable to include meaningful visual features as part of the localisation we abstracted camera images to a track limit representation.
The lack of visually salient features is discussed in Section \ref{sec:challengelimit}.
Unfortunately, Cartesian points provide much less information about match quality than re-observations of a know visual landmark.
This causes our localisation module to lack precision when approaching and travelling through corners, see Figure \ref{fig:localisation}(a,b).
If in the future visual features such as barriers, signs, grandstands, buildings, and trees are added revisiting SLAM algorithms would likely lead to significant improvements in localisation accuracy.

Due to this lack of accuracy we where unable to use the mapping and localisation module to drive from directly.
For these reasons our control is restricted to only the scene in which the perception module provides. 
In this challenge we do not have access to GPS localisation, inertial measurement unit data, or wheel speed data. 
This adds a considerable challenge for estimating if a given control input has produced the anticipated vehicle movement. 
The particle filter our localisation module relies on only uses knowledge of the observed track limits, control inputs, predicted behaviour, and that the vehicle has not left the track. 
For this reason we must have a considerable amount of uncertainty into the particle filter for it to converge and maintain localisation.

In its current form the MPC++ solution is quite fragile.
This is due to a combination of parameters that are required to be tuned and interplay with one another in unexpected ways.
One such example, discovered during experimentation for this work, is that the speed at which mapping is done has a significant impact on map quality and thus reference velocities, localisation, and consequently control.
We believe this can be addressed with map post-processing that would ensure consistency.
Many of the parameters used to configure localisation and the MPC solver have been manually tuned based on empirical observation.
It is not necessarily clear why these weights are the values which result in desirable behaviour and understanding the range and scale of values is often a process of trail and error.
Behaviour of the MPC solver is also not consistent under a uniform scaling of all term weights; indicting the optimisation problem is non-linear.
Due to this, we wish to formulate the tuning of these parameters as a safe learning problem or use the MPC as a control prior for a deep-controller.

Presently, MPC speed controls are not used once the vehicle has been localised.
Ideally, map reference velocities would be used to inform the MPC so that control inputs are predicted in advance. 
This ensures that the desired velocity is achieved at its associated track position, rather that an requested when the vehicle is at that position.
As an approximation to this desired behaviour, reference speeds given to the vehicle are from points in the map a fixed distance ahead of the vehicle's current position.
This ensures acceleration inputs are applied prior to a given point, enabling transition to the velocity required at the future point.

\subsection{Limitations of the challenge}\label{sec:challengelimit}
There where some aspects of how the challenge was setup that caused friction during the development of solutions we would like to address in the spirit of review and refinement.
Somewhat arbitrarily both receiving observations from the simulation and providing control input to the vehicle was limited to 10Hz.
Although it is understood that synchronisation of input and output is important we believe it would be better to leave this task in the hands of competitors.
Primarily our concern stems from the limitations around control input which prevent the use of granular inputs to make smoother changes to the vehicle's state.
By limiting input and observation to 10Hz each time control is set it is used by the vehicle for the next 100ms, preventing the use of a PID controller or other such method of smoothing inputs to the car.
This compounds when running on the prescribed evaluation server, which executes both simulation and solution code slower than our development machines, as if a control calculation takes longer than $100$ms to run we must wait until the next 100ms window to update observations and control.
Not only does this produce behaviour that rapidly jerks the vehicle, it does not provide sufficient time for control solutions to correct oversteer.
On this note, modern car safety systems detect and respond to loss of traction using on-board sensors. \cite{6928502}
When operating vehicles at racing pace it is crucial to control oversteer when cornering and apply threshold braking; both of which require an understanding of traction limits.
It is therefore hard to justify the challenge's decision to restrict access to this information, given its ultimate goals of safe racing.
The limited control over the camera's used on the vehicle also created tension that was somewhat remedied by the introduction of the multi-camera league.
We do understand that the positioning of the camera's maybe limited by the vehicle's design, but we see no reason that the resolution, rotation and lens type should be fixed.
By allowing camera parameter customisation specific solutions could tune cameras for specific goals.
For example, if the camera's goal is to observe track limits a developer might prioritise a wider field-of-view, minimise capturing vehicle bonnet and sky.
Some potential solutions where not able to be explored due to the nature of the simulation environment.

Initially we gravitated towards utilising simultaneous localisation and mapping (SLAM) algorithms for our localisation. 
We found that off the shelf monocular SLAM algorithms failed within the competition constraints.
This is likely due to the limited visual features, especially close to the vehicle, and limited field of view, refer to Figure \ref{fig:visualfeatures} and Figure \ref{fig:perception}.
This limitation effects both the diversity of potential solutions and a significant gap between solutions developed for the simulation versus those that are possible in reality.
Finally, we would like to see the criteria of exceeding the track limits expanded.
Currently vehicles may clip the white line with a single wheel and are considered to have committed a safety infraction.
Maximising track limits is a corner stone of racing enabling significant gains in lap time.
This practice decreases the cornering radius of turns enabling them to be taken with less lateral acceleration.
In our view this is ultimately safer then taking a different line with a tighter radius, requiring harsher control inputs.
Due to this we recommend the infraction be changed to accommodate the use of curbs on the track and allow for situations where two wheels maybe outside of white lines by some margin.

\subsection{Future work}
Our baseline can achieve competitive results on the supplied tracks under the simulation constraints. 
Much like our development thus far there are modules and improvements we would like to add in our quest to achieve optimal lap times.

Addressing the current limitations of our system would be first and foremost.
Reliable and Accurate track limit extraction is needed for better trajectory planning and increased agent confidence for upcoming control.
Mapping and localisation require suitably fine grained location information to enable control to follow a map based racing line, resulting in lower lap times.
Currently, the control system has no knowledge of grip and how it is traded between control inputs.
Building this knowledge into the system through vehicle dynamics would likely yield improved safety and speed.
Additionally, a learning based approach to MPC controller tuning would enable the automatic adaption of control parameters to unseen circuits.

Currently our system does not have a planning module which has a further look ahead than the MPC controller.
This kind of planning module could enable more efficient trajectories through corners, position the vehicle within the track limits to reduce cornering radius, and dynamically adjust control in response to vehicle state.

\section{Conclusion}
In this paper we have presented a modular autonomous control system for racing which can be incrementally improved on. 
The Learn To Race Challenge presented unique problems which prevented the application of previous autonomous racing solutions.
We have demonstrated the efficacy of a simple perception-mapping-localisation-control pipeline to control vehicles at racing pace safely.
Throughout the stages of this challenge we have built upon our previous solutions adding and improving modules. 
This approach has proved to be competitive with other agents submitted to the challenge.
In the future we would like to build additional machine learning components into our solution to see the agent safely push the vehicle to the limit and achieve optimal lap times. 

\section*{Acknowledgments}
This research was supported by Australian Government Research Training Program (RTP) Scholarships awarded to James Bockman and Matthew Howe.
Lockheed Martin Australia supported this research via a research scholarship awarded to James Bockman.

\bibliography{example_paper}
\bibliographystyle{icml2022}

\appendix
\input{appendix}

\end{document}

%% file: appendix.tex
\section{Modelling}\label{app:modelling}
All models where trained at an input resolution of $384x512$ in RGB colour space with no image augmentation.

\begin{figure}[h]
    \centering
    \resizebox{0.49\textwidth}{!}{
        \begin{tabular}{cc}
            \includegraphics[width=0.22\textwidth]{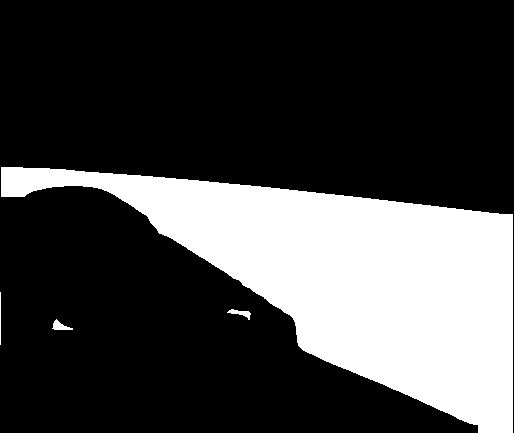}
             & \includegraphics[width=0.22\textwidth]{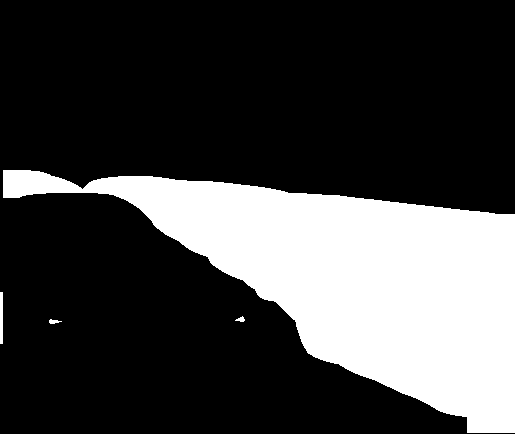}
        \end{tabular}
    }
    \caption{Example segmentation predictions from Deeplabv3+ using an output stride of $16$ (left) and $32$ (right)}
    \label{fig:void-suck}
\end{figure}

\subsection{EfficientNet Feature Pyramid Network}
The EfficinetNetV2 based FPN was trained using the Adam optimiser \cite{https://doi.org/10.48550/arxiv.1412.6980} and Dice loss \cite{DBLP:journals/corr/SudreLVOC17}.
A learning rate of $4$E-$4$ was used for $100$ epochs with a batch size of $2$.
The model with the lowest loss was selected for deployment.
We observed an average inference time on batches of $3$ images of $15.01$ ms on an NVIDIA RTX3090.

\subsection{ResNet Deeplabv3+}
Deeplabv3+ was trained using stochastic gradient decent (SGD) and cross-entropy loss.
SDG was configure to use a learning rate of $0.03$, momentum of $0.9$ and weight decay of $5$E-$5$.
A learning rate of $0.003$ was used for $30$ epochs with a batch size of $8$.
The learning rate was reduced by a factor of $0.1$ each time the validation loss did not change more than $1$E-$4$ for $3$ epochs.
The model weights present after the full number of epochs was selected for deployment.
As all variants explored fit data extremely well--with average IOU scores larger than 98\%---ResNet-18 with an output stride of $16$ was selected as a trade off between inference time and mask quality, see Table \ref{tab:deeplab-inference-time}.
Although, an output stride of $32$ would have been faster, qualitative inspection of the masks found that the large amount of up-sampling required caused over smoothing in masks that would clip into the car's silhouette, see Figure \ref{fig:void-suck}.
We observed an average inference time on batches of $3$ images of $5.28$ ms on an NVIDIA RTX3090.

\begin{table}[h]
\begin{tabular}{@{}lcccc@{}}
\toprule
              & \multicolumn{4}{c}{\textbf{ResNet Variant Inference Time (ms)}}        \\ 
\textbf{Output Stride} & 18       & 32       & 50       & 101      \\\midrule
8             & 13.08  & 22.78  & 41.97  & 64.83  \\
16            & 5.28   & 7.4    & 15.83  & 21.98  \\
32            & 3.74   & 5.6    & 11.46  & 18.01  \\ \bottomrule
\end{tabular}
\caption{Average inference times observed on batches of 3 images for Deeplabv3+ models}
\label{tab:deeplab-inference-time}
\end{table}